\documentclass[conference]{IEEEtran}
\usepackage{cite}
\usepackage{amsmath, amssymb}
\usepackage{graphicx}
\usepackage{hyperref}
\usepackage{url}
\usepackage{xcolor}
\usepackage{booktabs}
\usepackage{multirow}
\usepackage{listings}
\usepackage{enumitem}
\usepackage{caption}
\usepackage{algorithm}
\usepackage{algpseudocode}
\usepackage{microtype}   

\usepackage{longtable}
\hypersetup{
    colorlinks=true,
    urlcolor=blue,
    citecolor=blue,
    linkcolor=blue
}
\usepackage{subcaption}
\usepackage{soul}

\lstdefinelanguage{Gherkin}{
  morekeywords={Feature,Scenario,Given,When,Then,And,But},
  sensitive=false,
  morecomment=[l]{\#},
  morestring=[b]",
}
\usepackage{comment}
\title{
Multi-Agent LLM-based Metamorphic Testing for REST APIs
}

\author{
\IEEEauthorblockN{
Shehroz Khan,
Abdullah Mughees,
Gaadha Sudheerbabu,
Tanwir Ahmad,
Dragos Truscan}
\IEEEauthorblockA{
Åbo Akademi University\\
Turku, Finland\\
\texttt{firstname.lastname@abo.fi}
}
}

\begin{document}
\maketitle
\begin{abstract}
As REST APIs become an increasingly significant part of software systems, their validation is becoming more critical. Hence, testing and uncovering underlying issues are of utmost importance for improving software quality. However, testing REST APIs is challenging mainly due to the difficulty of assessing whether the output of an API call is correct, i.e., the test oracle problem. Metamorphic testing is a specification-based testing approach for situations where correct outputs are unknown or not specified explicitly. To check the correctness of a system, relations between the different outputs are specified.
We present \textbf{ARMeta}, a tool-supported approach that uses an LLM-based multi-agent workflow to support metamorphic testing of REST APIs documented with OpenAPI. The agentic workflow is used to identify metamorphic test scenarios and specify them in the Given--When--Then format. 
These scenarios are automatically implemented as executable 
tests and executed against the system under test.

We evaluate ARMeta on two publicly available web applications that expose REST interfaces and compare its performance with a scenario-based testing baseline.
The results show that ARMeta explores behaviors that serve as a complement to existing scenario-based testing approaches.
\end{abstract}


\section{Introduction}\label{sec:introduction}

REST APIs are used in many software systems, but they are not easy to test well. The OpenAPI Specification is commonly used to document REST API services, including older specifications that use the Swagger 2.0 format. These specifications describe endpoints, methods, parameters, and data models, which makes them a practical input for automated test generation~\cite{casas2021uses}.  However, such specifications often do not give a clear expected output for each request. Outputs can change because the system state changes, the database changes, or other clients are using the API at the same time. For this reason, many automated tests only check status codes and response schemas. These checks help, but they do not test deeper behaviour.


Metamorphic testing (MT) is a specification-based testing technique that can be used to test systems lacking explicit test oracles~\cite{chen2020metamorphic}. MT checks whether multiple executions of the system under test satisfy specific necessary properties, called \textit{metamorphic relations} (MRs). It starts with a \textit{seed input} and derives one or more \textit{follow-up inputs} by applying a \textit{metamorphic transformation}. Instead of checking one output against a fixed expected value, the test verdict is assigned based on whether the MR that links the seed and follow-up outcomes holds 
\cite{liu2012new}. MT has been applied in many application domains, including web services~\cite{sun2011metamorphic, segura2018metamorphic}. Open challenges remain in systematically identifying effective MRs and reducing reliance on domain experts~\cite{chen2018metamorphic}.

Despite extensive work on REST API testing and prior studies on metamorphic testing, there is still no end-to-end, specification-driven approach that systematically applies metamorphic testing to REST APIs in an executable and automated manner. Existing specification-based API testing tools mainly focus on individual requests and basic checks such as status codes or schema conformance, while scenario-based approaches rely on concrete examples and expected outcomes. Although metamorphic testing can alleviate the test oracle problem, practical challenges remain in identifying suitable metamorphic relations, grounding them in API specifications, and operationalizing them as executable tests for real REST APIs. Moreover, prior work~\cite{Segura2016Survey} acknowledges that follow-up executions required for metamorphic testing may fail due to specification–implementation mismatches or robustness issues, yet current approaches do not systematically generate, execute, and report such transformation-based tests and their outcomes. This gap motivates the need for an automated workflow that derives metamorphic test scenarios from OpenAPI specifications, executes them against real APIs, and reports both relation-level and execution-level outcomes as meaningful testing results.

Large language models (LLMs) are increasingly used for automated test generation, and recent studies~\cite{luu2023chatgpt, zhang2023automated, shin2024generatingexecutablemetamorphicrelations} show that LLMs can identify MRs in different application domains, such as web applications, autonomous driving systems, and embedded systems. Although these studies demonstrate the capability of LLMs to generate MRs, they also highlight that not all identified MRs are reliable by default and require guardrails to ensure a high level of accuracy. However, existing work does not explain how metamorphic relations can be systematically derived from REST API specifications and transformed into specification-grounded, executable tests while coping with execution failures and contract mismatches. Two main problems remain:

\begin{enumerate}[leftmargin=*]
\item Metamorphic-style transformations are not inferred directly from REST API specifications in a way that yields clear seed--follow-up tests grounded in documented operations.
\item There is no end-to-end approach that transforms these transformation-driven scenarios into implemented and executable REST API tests while handling LLM instability and specification misalignment.
\end{enumerate}

In our work, we investigate the following research questions: 
\begin{itemize}
    \item \textit{RQ1 (Metamorphic test scenario generation):} How well can an LLM generate candidate metamorphic test scenarios? 
    \item \textit{RQ2 (Executable metamorphic tests):} how effectively can a multi-agent workflow turn these candidate metamorphic test scenarios into executable metamorphic tests that reveal faults?
    \item \textit{RQ3 (Complementarity):} to what extent does ARMeta 
    complement existing scenario-based API testing approaches?
\end{itemize}

To this extent,  we introduce \textbf{ARMeta}, a tool-supported approach that uses a multi-agent workflow to support metamorphic testing for REST APIs using the API specification as the main input. ARMeta generates MT scenarios in a Given-When-Then (GWT) format, links them to the OpenAPI specification, and converts them into executable tests. The tests are implemented as Gherkin feature files and Python step definitions using the \texttt{Behave} \cite{behave-pypi} package. ARMeta executes the tests and produces a report of the results. We provide ARMeta as a Streamlit GUI ~\cite{streamlit_docs} so users can execute the pipeline, view the generated artifacts, and inspect failures. To summarize, the contributions of this paper are as follows:
\begin{itemize}[leftmargin=*]
    \item LLMs are used to identify and extract metamorphic relations from REST API specifications;
    \item Metamorphic relations are specified as GWT requirements patterns, allowing for better processing by LLMs and for easier inspection by human-in-the-loop;
    \item The end-to-end workflow uses a multi-agent approach to extract metamorphic relations and then generate, evaluate, refine, and select executable metamorphic tests from a high-level metamorphic testing specification, finally reporting the test results.
    \item The workflow monitors operational coverage of the API and decides to generate additional metamorphic tests if needed or to stop if a certain time budget is reached.
\end{itemize}

The rest of this paper is organized as follows. Section~II discusses related work. Section~III presents the ARMeta workflow and its main components. Section~IV describes the experimental design and reports the evaluation results. Section~V discusses the limitations and outlines future work. Section~VI concludes the paper.


\section{Related Work}
\label{sec:related_work}

Many existing tools generate REST API tests from the specification and execute them automatically. Their checks commonly focus on unexpected status codes, server crashes, and schema mismatches. These tools are effective for finding robustness problems, but they usually do not encode higher-level properties as explicit test oracles \cite{golmohammadi2023testing}.

Metamorphic testing has also been applied to web services and RESTful APIs. For example, Sun et al.~\cite{sun2011metamorphic} and Segura et al.~\cite{segura2018metamorphic} study metamorphic relations for service/API behaviors and how to apply them to service calls. Our work differs in that ARMeta targets an \emph{end-to-end, specification-driven} workflow: it infers and refines candidate relations from OpenAPI text, expresses them as GWT Metamorphic test scenarios over seed and follow-up executions, and synthesizes and executes runnable tests automatically, with explicit checks for specification alignment and successful execution. 

Recent work uses large language models for REST API testing. LogiAgent~\cite{zhang2025logiagent}, for example, proposes a multi-agent pipeline that generates scenarios, sends requests, and validates responses using expectations derived from the documentation and the scenario context. This supports more semantic checking than schema-only validation, but the decision logic is still based on whether the response appears consistent with the scenario rather than on a strict relation between seed and follow-up executions. 
RESTifAI focuses on generating reusable test suites by first producing a valid workflow and then extending it with additional tests, including negative cases, with emphasis on rerun and integration into development pipelines \cite{kogler2025restifai}. 

AutoMT~\cite{automt2025} is a multi-agent tool that applies metamorphic testing to autonomous driving systems. It derives MRs from driving rules, creates follow-up tests from existing tests, executes them, and reports violations~\cite{automt2025}. AutoMT is not designed for REST API testing and depends on the driving domain and execution setup. In contrast, our approach targets REST API services described by OpenAPI/Swagger. We write each test as a Metamorphic test scenario with a seed execution and a transformed follow-up execution derived from inferred MR checks.

One study \cite{shin2024generatingexecutablemetamorphicrelations} presents a method that uses large language models to derive metamorphic relations from natural-language requirements and convert them into executable forms using a domain-specific language called SMRL. Their work shows that LLMs can help identify transformation-based test relations and reduce manual effort in metamorphic testing. Although API specifications are provided when converting relations into executable form, the relations themselves are mainly derived from textual requirements and are not systematically organized around individual REST API operations. In contrast, our approach treats the OpenAPI specification as the primary source for generating and validating metamorphic tests. specifications such as OpenAPI. It does not directly derive clear seed–follow-up test scenarios from documented API operations. In addition, the paper does not provide a complete end-to-end solution that generates, executes, and validates REST API tests automatically, nor does it fully address issues such as LLM instability or mismatches between generated tests and API specifications. Therefore, while the work is a useful step toward LLM-assisted metamorphic testing, important challenges remain for practical, specification-driven REST API testing.

Our approach differs mainly in how it structures tests and how it interprets outcomes. We use the OpenAPI Specification as a strict source of truth for selecting valid endpoints and parameters, and we encode each Metamorphic test scenario with an explicit seed and transformed follow-up. When a relation assertion is reachable and violated, the outcome is an assertion failure. However, many outcomes are execution-level or contract-level failures that occur before reaching a relation assertion (e.g., HTTP 5xx, exceptions, or undocumented status codes); we report these as scenario failures because they represent robustness/contract signals triggered along the transformed follow-up path. We also compute operation coverage to steer generation and to sanity-check which documented operations are being exercised; this coverage is approximate and not used as a head-to-head comparison score.

We also address practical instability in large language model outputs by separating generation and refinement. A stable model produces initial relations and test code, and a stronger model refines the code to better match the specification and reduce unsafe patterns. This can improve repeatability, but full determinism is still difficult in large language model-based pipelines \cite{atil2024non}.

\section{Overview of the Approach}
\label{sec:approach}

\subsection{Conceptual approach}

In our approach, an MR is defined as a relation across two executions of an API: a seed execution and a follow-up execution. The seed input is the baseline API call (or a short sequence of calls), and the seed output is what we observe from it, such as a response body, a list size, or a key field value. The follow-up input is built by transforming the seed input or the API state, and the follow-up output is the response we observe after that change. The MR oracle is not a single expected output; it is the relationship between the two observed outputs (i.e., HTTP responses).

In the first step of our approach, we extract \textit{high-level metamorphic tests} (HLMT) from the API specification, and we express them with Gherkin's language Given-When-Then (GWT) requirements patterns~\cite{wynne2012cucumber}. This allows for each metamorphic test to be expressed in a structured, readable form that both humans and machines can more easily process. An example is shown in Figure \ref{fig:mr-petstore-subfig}-(a). The \textit{Scenario}  describes the high-level goal of the metamorphic test. The \textit{Given} step will describe the seed input, and the \textit{When} step will describe how the follow-up input is derived from the seed input by applying a metamorphic transformation. Each of these steps corresponds to one or several API requests, each of which is expected to return a response containing a JSON payload (see lines 19 and 36 in Figure \ref{fig:mr-petstore-subfig}-(b)). Such requests include modifying a single field in the request payload or resource representation, creating or deleting a resource, and applying a filter to the retrieved data. The \textit{Then} step states the expected relationship, i.e., the MR, between follow-up output and seed output, using relations such as equality, inclusion, exclusion, or difference. 

Subsequently, each HLMT is translated into an \textit{executable metamorphic test} (EMT) by mapping HLMTs to the REST API specification. For convenience, we use the Python \textit{behave framework}~\cite{behave_docs2026}, as implementation language, since it directly supports Gherkin syntax and allows these metamorphic tests to be automatically executed once the corresponding step definitions have been implemented. In \textit{behave}, the GWT steps are specified as functions with corresponding decorators (see Figure \ref{fig:mr-petstore-subfig}-(b)). In our approach, we ensure that in order to check the MR defined between seed and follow-up outputs in the \textit{Then} step, both the \textit{Given} and \textit{When} steps must execute successfully to collect the outputs. Therefore, the generated Behave code may include assertion statements or raise exceptions if the HTTP response indicates an error (e.g., lines 17, 29, and 35) in the \textit{Then} step for the MR check between the outputs generated in the \textit{Given} and \textit{When} steps.


\begin{figure}[t]
\centering

\begin{subfigure}{\columnwidth}
\centering
\ttfamily\tiny\raggedright
{\setlength{\parindent}{0pt}
\textbf{Scenario:} Updating a pet's status should be observable when retrieving that pet by its identifier.\par
\hspace{1pc}\textbf{Given} a seed input that retrieves a pet by its identifier using the pet-retrieval-by-identifier operation, producing a seed output with the pet's current status.\par
\hspace{1pc}\textbf{When} a follow-up input is created by updating the pet's status using the pet-update-with-form-data operation and then retrieving the same pet by the same identifier again using the pet-retrieval-by-identifier operation, yielding a follow-up output.\par
\hspace{1pc}\textbf{Then} the follow-up output should reflect the updated status for that pet, and the pet's identifier should remain the same as in the seed output.
}
\caption{}
\end{subfigure}

\vspace{0.4em}

\lstset{
  language=Python,
  basicstyle=\ttfamily\tiny,
  columns=fullflexible,
  keepspaces=true,
  breaklines=true,
  showstringspaces=false,
  frame=single,
  numbers=left,
  xleftmargin=4em,
  framexleftmargin=3em,
  emph={%
    @given, @when, @then%
    },emphstyle={\bfseries}%
}%

\begin{subfigure}{\columnwidth}
\centering
\begin{lstlisting}
from behave import given, when,then
import requests

@given("a seed input that retrieves a pet by its identifier using the pet-retrieval-by-identifier operation, producing a seed output with the pet's current status for MR26.")
def _mr_MR26_seed_input(context):
    pet_data = {
        "id": 1,
        "name": "doggie",
        "photoUrls": ["url1"],
        "status": "available",
    }
    
    create_resp = requests.post(f"{BASE_URL}/pet", json=pet_data, timeout=10)
    create_resp.raise_for_status()
    
    seed_resp = requests.get(f"{BASE_URL}/pet/{pet_data['id']}", timeout=10)
    seed_resp.raise_for_status()

    context.seed_output = seed_resp.json()

@when("a follow-up input is created by updating the pet's status using the pet-update-with-form-data operation and then retrieving the same pet by the same identifier again using the pet-retrieval-by-identifier operation, yielding a follow-up output for MR26.")
def _mr_MR26_followup_input(context):
    updated_status = "sold"

    post_resp=requests.post(
        f"{BASE_URL}/pet/{context.seed_output['id']}?status={updated_status}",
        timeout=10
    )
    post_resp.raise_for_status()

    follow_resp = requests.get(
        f"{BASE_URL}/pet/{context.seed_output['id']}",
        timeout=10
    )
    follow_resp.raise_for_status()
    context.followup_output = follow_resp.json()


@then("the follow-up output should reflect the updated status for that pet, and the pet's identifier should remain the same as in the seed output for MR26.")
def _mr_MR26_followup_output(context):
    seed = context.seed_output
    follow = context.followup_output

    # follow-up output should be identical to seed output
    # except for the 'status' field, which must be updated
    assert follow["id"] == seed["id"]
    assert follow["status"] != seed["status"]
\end{lstlisting}
\caption{}
\end{subfigure}
\caption{Example of High-level MT scenario in GWT form (a) and of Executable MT scenario implementation (b)}
\label{fig:mr-petstore-subfig}
\end{figure}

\subsection{Multi-agent workflow}
\label{sec:agentic_workflow}

\begin{figure}[t] \centering \includegraphics[width=.5\textwidth, trim={1cm 0cm 0cm 0cm}, clip]{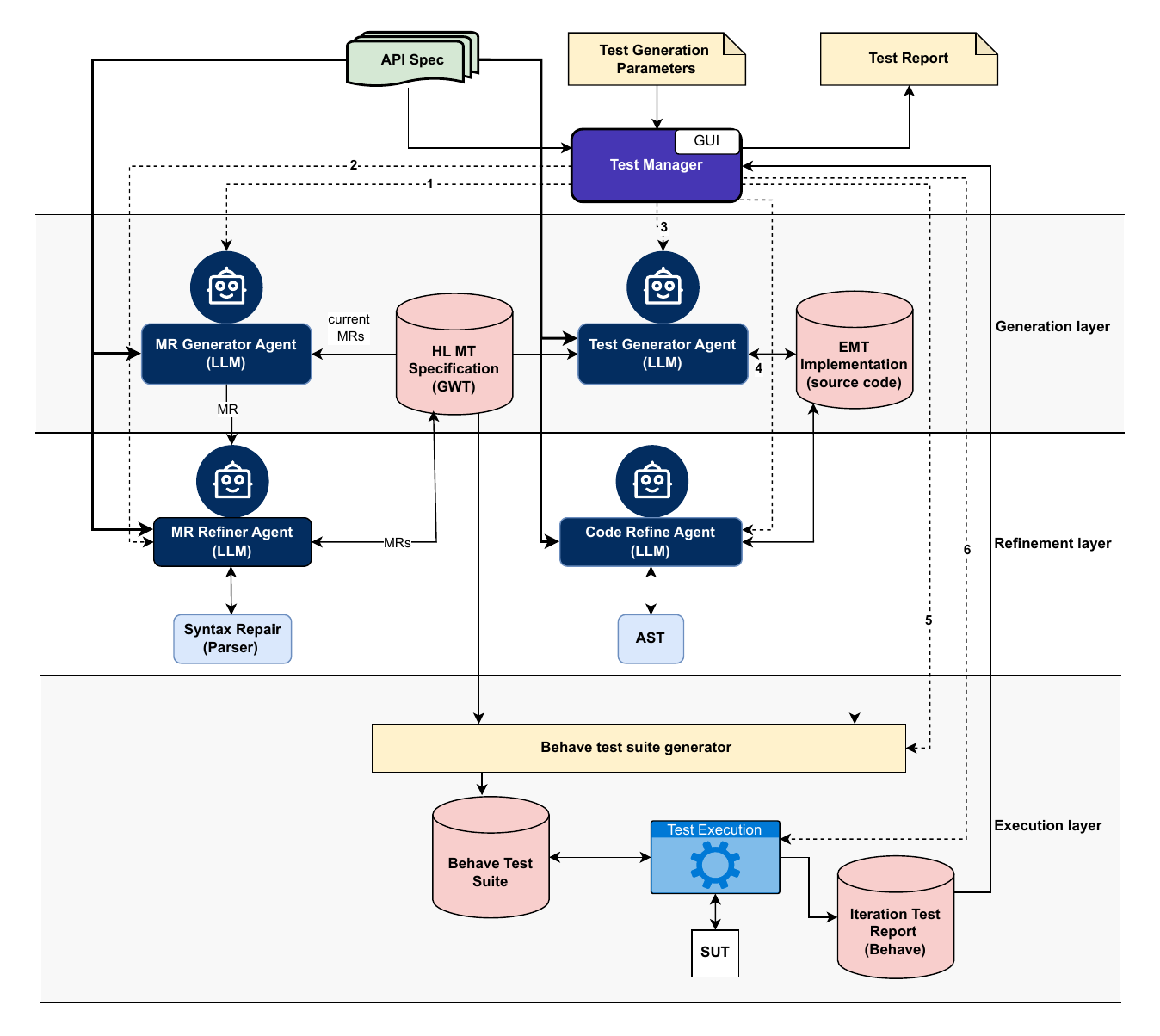} \caption{High-level multi-agent architecture} \label{fig:architecture} \end{figure}

The agentic workflow implementing our approach is illustrated in Figure \ref{fig:architecture}.  
The pipeline is controlled by a \textit{Test Manager} and several LLM-based agents with clearly defined roles. The approach is iterative and incremental. A test generation session is composed of several iterations. In each iteration, a set of metamorphic tests is generated and executed. If the stopping criteria for test generation are not met, a subsequent iteration is initiated. 

Concretely, the \textit{Test Manager} takes the REST API specification and the test generation parameters as input. The latter are defined as follows:
\begin{itemize}
\item{Agent configuration}
    \begin{itemize}
    \item \textit{Model selection}: specifies the LLM used at each stage of the pipeline
    , with models assigned to agents based on capability and cost.
    \item \textit{Model temperature}: controls randomness in test generation. Each LLM instance maintains an independent temperature setting via the \texttt{temperature} field in the LLM API call.
    \end{itemize}
    
\item{Stopping criteria}
    \begin{itemize}
    \item \textit{Target operational coverage}: specifies the percentage of API operations, defined as unique combinations of endpoints and HTTP methods, that must be exercised by the generated test suite.
    \item \textit{Plateau window}: defines the number of consecutive iterations allowed without discovering new tests or an increase in the operational coverage. 
    \item \textit{Request budget}: sets an upper bound on the total number of API requests permitted during a single session. 
    \item \textit{Time budget}: specifies the maximum wall-clock duration of a single session, after which execution is halted. 
    \end{itemize}
    \end{itemize}


In each iteration, the \textit{MR Generator Agent} proposes a configurable bounded batch of new HLMT candidates. The prompt explicitly requires the candidates to be different from those generated in previous iterations by providing the agent with the accumulated HLMT history. LLM responses are converted into a syntactically correct JSON list so that subsequent stages can reliably iterate over individual HLMT objects. This normalization is performed using a lightweight parser that tolerates minor formatting deviations, such as JSON-like or Python-literal formatting, repairs small syntactic issues when possible, and extracts the HLMT objects into a consistent list structure. Next, an \textit{MR Refiner Agent} aligns the selected HLMTs with the OpenAPI specification by checking their semantic consistency with the documented API behavior. Relations that cannot be supported by the specification are dropped, and the remaining ones form the HLMT specification for the iteration.

 For each HLMT, the \textit{Test Generator Agent} generates Behave step definitions whose decorators match the HLMT Given-When-Then strings exactly. A \textit{Code Refiner Agent} validates the generated step code using an abstract syntax tree (AST) check using Python’s \texttt{ast.parse()} function to ensure syntactic correctness. If the AST validation fails, a bounded repair loop is triggered in which the agent is asked to repair the syntax while keeping the step decorators unchanged to preserve binding to the HLMT specification. The repair loop is limited to a configurable number of attempts as defined in the pipeline configuration.
 If a valid step code still cannot be produced after these attempts, the pipeline inserts minimal placeholder step definitions. These placeholders are syntactically valid no-op steps that do not perform API interactions to ensure mapping with the HLMT and that the Behave runner can still load and execute the remaining scenarios. Although such cases are marked as failures during execution, they have been discarded during test execution. 
 
After generating the EMTs, the pipeline assembles each iteration into a Behave-compatible test suite.  The resulting directory structure follows the conventional Behave layout, where a feature file (e.g., \texttt{iteration\_01.feature}) contains multiple scenarios, one per HLMT, and the corresponding step definitions are implemented in a shared module (e.g., \texttt{features/steps/iteration\_01\_steps.py}). This organization preserves the logical grouping of HLMTs by iteration while ensuring full compatibility with the Behave runner. The suite is executed using Python’s \texttt{behave} runner. After execution, the pipeline collects passed and failed scenarios and error traces, updates operation coverage and other metrics, and evaluates whether the configured stopping criteria are met. If and only if the stop criteria are satisfied, the pipeline saves the session artifacts, including the HLMT specifications, generated tests, execution logs, and coverage summaries, and produces a consolidated session report. Otherwise, a new iteration is started.

\subsection{Implementation}


ARMeta is implemented in Python, as three layers: generation, refinement, and execution layers. 
This separation helps with debugging and evaluation, since errors can be traced to a specific layer, and individual components can be modified or replaced without affecting the rest of the system. The layered design also allows different LLM model settings to be used at different stages, enabling more flexible reasoning early on and more stable behavior during execution. In particular, different configurable models can be used at each layer of the pipeline. Overall, this structure makes ARMeta easier to extend and adapt to new tasks.

The \textit{generation layer} is implemented using CrewAI~\cite{CrewAI}, which provides convenient abstractions for defining agents, as well as good support for lower models in which the temperature can be controlled. 
In CrewAI, the LLM-driven components are structured as \textit{agents} and \textit{tasks}. Listing~\ref{lst:crewai_agent_task} shows an excerpt of the MR Generator Agent. Each agent is defined in CrewAI by a role, a goal, a backstory, and an LLM. The \textit{role} specifies the agent's responsibility (e.g., relation generation or code refinement), the \textit{goal} constrains the expected output, and the \textit{backstory} provides an instruction prompt with domain guidance and output constraints. Each agent is bound to a configured \textit{llm} interface that specifies the provider/model and decoding parameters; no model training is performed.  
For instance, the prompt shown in Listing~1 accepts the following parameters: \{openapi\_spec\}, \{no\_tests\}, \{base\_url\}, and \{prev\_tests\}. Here, \{openapi\_spec\} denotes the OpenAPI specification of the target service, \{no\_tests\} specifies the number of HLMTs to be generated, and \{base\_url\} represents the endpoint of the SUT. Test generation is performed iteratively in batches, at iteration $i$, \{prev\_tests\} includes the complete set of HLMTs produced in iterations $1$ through $i-1$, which are provided as contextual input to guide subsequent test generation. To enforce output format, we include a single illustrative JSON example in the prompt (e.g., retrieving a list of resources and comparing a count). This helps anchor the required structure but may mildly bias the surface wording toward the example domain.

Work units are represented as \textit{tasks} that contain the \textit{description}, the \textit{expected output}, and inputs such as the API specification, base URL, and previous metamorphic tests. An agent can be linked to several tasks. 

The \textit{refinement layer} is implemented outside of CrewAI using direct LLM API calls. This design choice avoids the model limitations of the CrewAI interface and allows us to use a more capable model for refinement. In the implementation, refinement is realized as two standalone functions that the Test Manager calls at fixed points in the pipeline: one refines the HLMT list after generation, and the other refines the step-definition code after code generation. Both functions take the OpenAPI specification as an explicit input and return a revised artifact of the same type, i.e., a JSON HLMT list in the first case and a Python step module in the second case.

\begin{lstlisting}[language=Python,
    frame=single,
    basicstyle=\ttfamily\tiny,
    numbers=left,
    numberstyle=\tiny,
    xleftmargin=4em,
    framexleftmargin=3em,
    caption={CrewAI agent and task definition for HLMT generation (prompt excerpt).}, label={lst:crewai_agent_task}]
mr_llm = build_llm(provider, model, api_key, base_url, temperature, seed, no_tests)
...
mr_agent = Agent(
    role="MR Generator",
    goal="Generate up to {no_tests} unique property-based Metamorphic Relations using Given/When/Then ...",
    backstory="Metamorphic testing verifies relationships between multiple executions of a system ...",
    llm=mr_llm,
)

mr_task = Task(
    description="You are a Metamorphic Relation Generator. Metamorphic testing verifies relationships as per open api specification {openapi_spec}.. with base url {base_url}...do not include previous generated metamorphic tests {prev_tests}...",
    expected_output="ONLY a JSON array of objects with fields id scenario given when then ...",
    agent=mr_agent,
)
...
Crew([mr_agent], [mr_task]).kickoff(inputs)
\end{lstlisting}

The \textit{execution layer} handles the execution of tests for each iteration using the behave framework and collecting a report that summarizes the execution results, indicating which scenarios have passed or failed. For each scenario, the report shows the status of its individual GWT steps. A scenario 
is considered passed only if all its steps pass; if any step fails, the entire scenario is marked as failed. After each iteration, the \textit{Test Manager} parses the iteration-level execution report to extract the pass/fail outcome and error details for each scenario and stores an iteration summary. When the session terminates, the \textit{Test Manager} aggregates the iteration summaries across all iterations into a single session-level summary that consolidates the overall pass/fail outcomes and execution statistics.

\begin{figure}[t]
    \centering
    \includegraphics[width=0.95\linewidth]{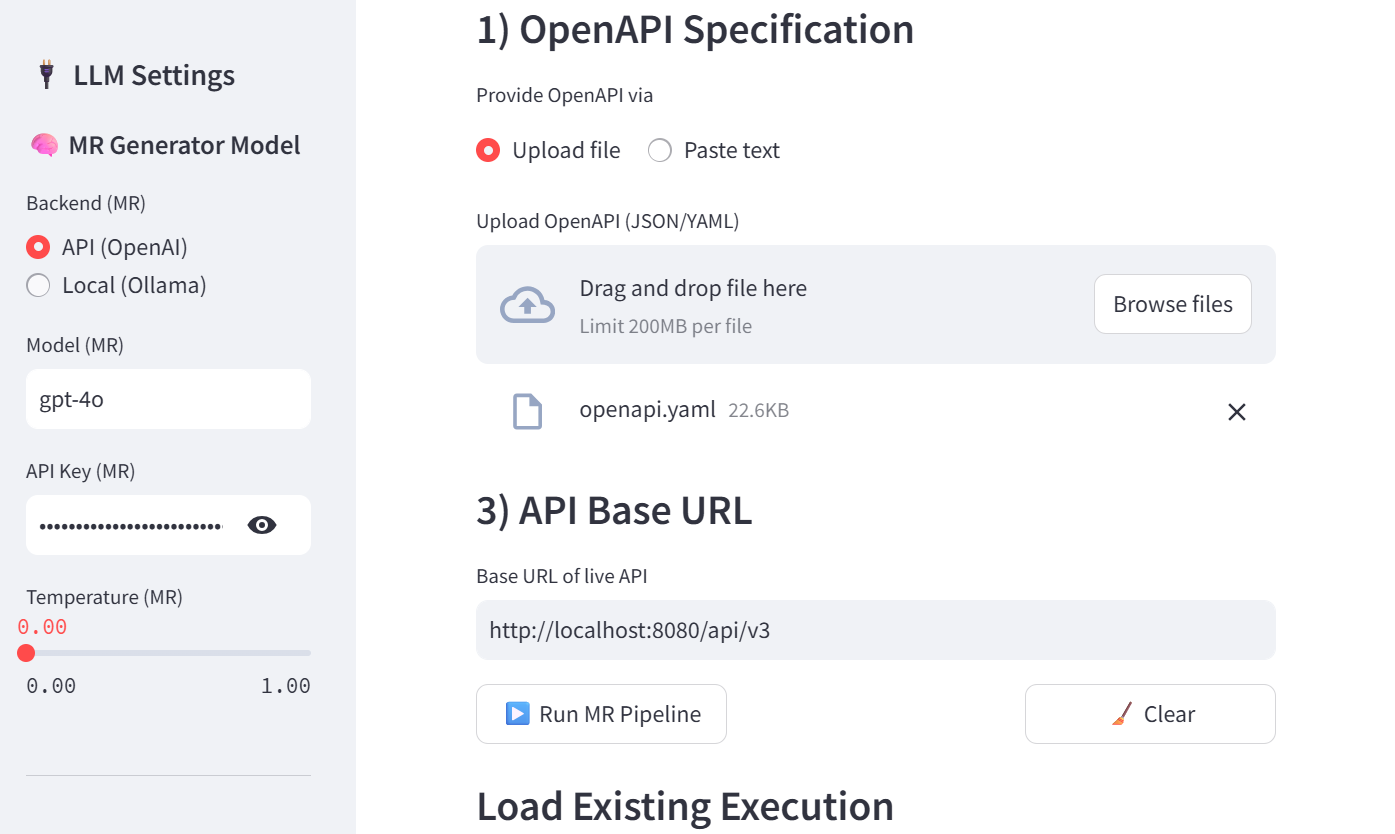}
    \hfill
    \includegraphics[width=0.95\linewidth]{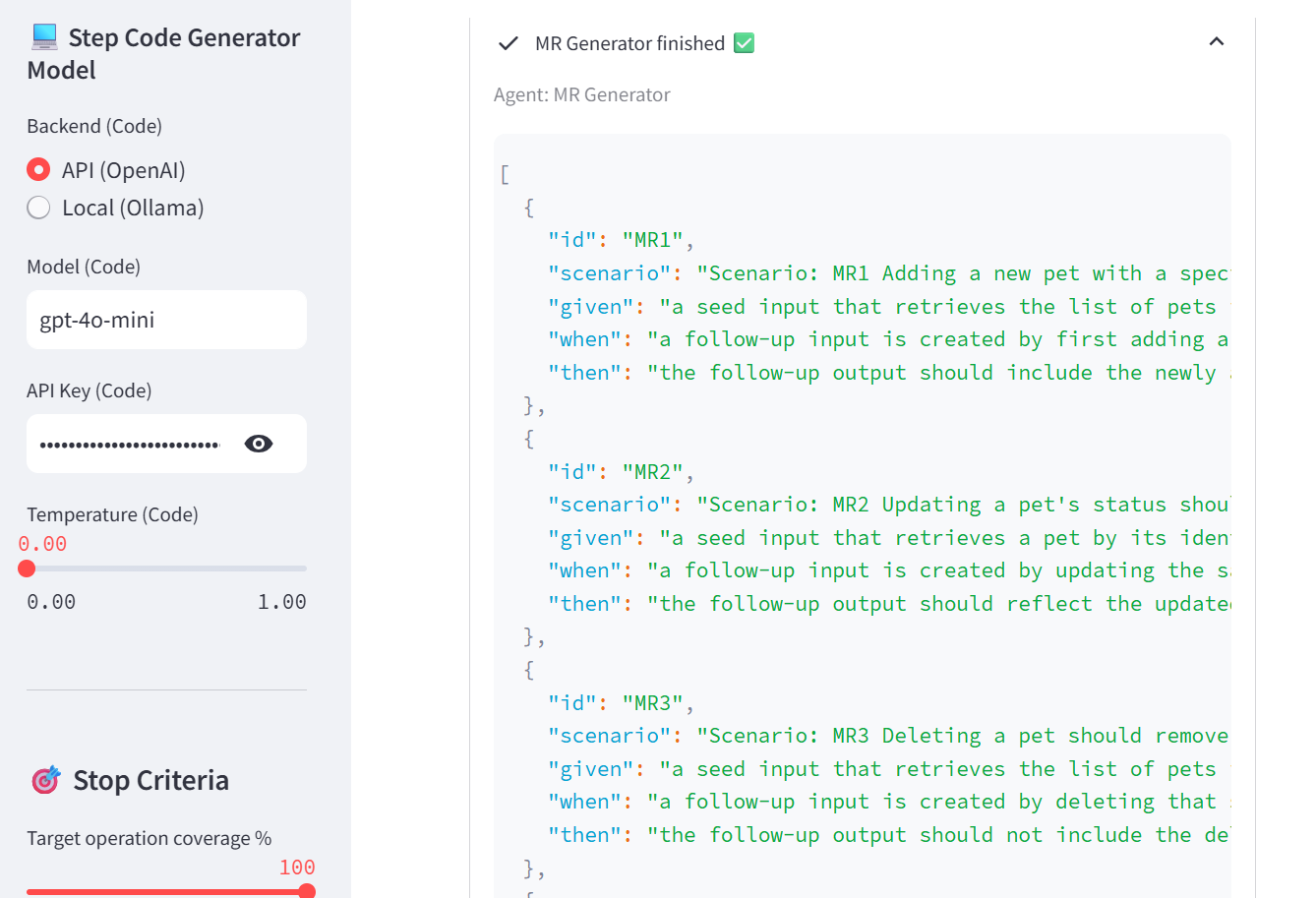}
\hfill
    \includegraphics[width=0.95\linewidth]{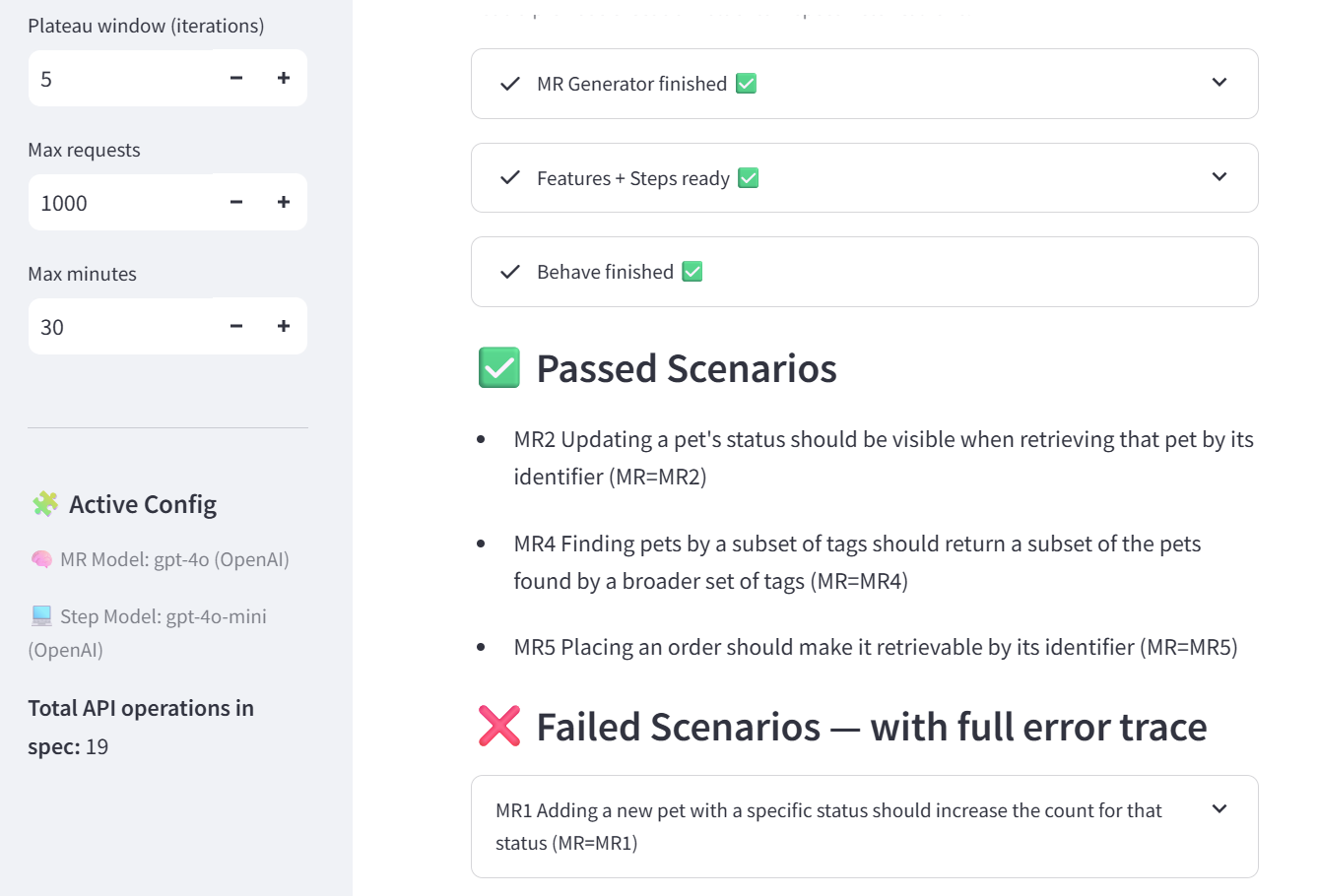}
    \caption{ARMeta tool displaying the passed and failed scenarios.}
    \label{fig:armeta}
\end{figure}

A GUI is implemented using Streamlit \cite{Streamlit}, that collects the API specification and SUT base URL, exposes key configuration parameters, executes the pipeline, and visualises the generated artefacts and execution reports. Figure~\ref{fig:armeta} shows the ARMeta graphical user interface, including the visualization of passed and failed scenarios.

\section{Evaluation}
\label{sec:evaluation}
This section evaluates ARMeta in order to provide an answer to the research questions discussed in Section \ref{sec:introduction}. We split the evaluation into three parts: test generation, test execution, and comparison with the LogiAgent approach, which was discussed in the related work. LogiAgent was chosen as a reference because it uses a multi-agent system targeted at REST APIs, without focusing on metamorphic tests.

\subsection{Case Studies}
For evaluation, we use two publicly available REST API case studies.  \textbf{PetStore} \cite{petstore} is a REST service with an OpenAPI specification describing 19 operations~\cite{petstore}. An operation is an endpoint path and the associated method.
\textbf{UserManagement}~\cite{usermanagement} is a REST service with a legacy Swagger 2.0 specification describing 28 operations, and it includes role-based access control (RBAC)~\cite{usermanagement}. We selected these two systems because they are commonly used as benchmarks in prior REST API testing work ~\cite{zhang2025logiagent}~\cite{ kim2022automated}.

Each case study and tool was deployed locally on a Windows-based PC (13th Gen Intel Core i7-1360P, 2.20~GHz, 32~GB RAM, 4~GB dedicated graphics card, 954~GB storage). 

\subsection{Workflow configuration}
For our experiments, we configure the workflow as follows:

\begin{itemize}
\item \textit{Model selection}: GPT-4o is used for MR generator agent, GPT-4o-mini for code generator agent, and GPT-5.1 is used for both refiner agents.

\item \textit{Model temperature}: Set to 0 to reduce randomness in the model’s output. Although the model may still show small variations, this setting makes the results more consistent. This is important for reliable test scenario generation, test code generation, and refinement.
\item \textit{Target operational coverage}: 100\%
\item \textit{Plateau window}: 5 iterations is based on the time budget, and after performing experiments with lower and higher plateau windows.
\item \textit{Request budget}: 1000 requests.
\item \textit{Repair loop count}: 2.
\item \textit{Time budget}:  30 minutes.
\end{itemize}

\subsection{Evaluation metrics}
\label{sec:experimental_design}

We evaluate ARMeta using the following metrics: 
\emph{Generated HLMTs} denote the number of high-level metamorphic test scenarios produced per test session. 
\emph{Generated EMTs} is the number of executable metamorphic tests derived from HLMTs and successfully executed (e.g., without any syntax errors) against the SUT.
An EMT is considered \emph{passed} when all its Given--When--Then steps, including the metamorphic relation assertion, execute successfully; it is considered \emph{failed} when any step fails due to an assertion failure or an HTTP/runtime error.
\emph{Operational coverage} measures the percentage of OpenAPI operations (unique endpoint--method pairs) exercised by the executed EMTs.
\emph{False positives failing tests (FPFT)} are failures caused by incorrect test construction or incorrect assumptions derived from the specification rather than by faults in the SUT; these are identified through manual inspection of execution traces and logs. The remaining failing tests reveal faults in the SUT, and are considered \emph{true positives failing tests (TPFT)}. We also define the percentage of TPFTs among failed tests as \emph{true positive rate (TPR)}. 
To benchmark the variation among HLMT generated in each session,  we define a metric, \textit{semantically unique} HLMT, to identify the tests that have the same intent but are specified or implemented in a slightly different way. 
Semantically equivalent HLMT are computed by processing all HLMT scenario titles from every session and normalizing them into a deterministic \texttt{semantic\_group\_slug}. Normalization standardizes titles into a comparable form (e.g., casing, punctuation, ID placeholders), and grouping simply collects all titles that map to the same slug as one semantic group.

Specifically, for each scenario title, we apply a lightweight text normalization pipeline: we convert the text to lowercase, remove any explicit MR identifiers (e.g., ``MR~12'') that may be appended by the generator, replace non-alphanumeric character spans with underscores, and collapse repeated underscores. The resulting slug is used as a stable key that is robust to minor surface variations in phrasing while preserving the underlying metamorphic intent. Finally, we group together all scenario instances that map to the same slug across all sessions and iterations, and we report the number of distinct slugs as the number of semantically unique metamorphic scenarios. For example, the semantic group slug:

 \begin{center}
 {\scriptsize
 \texttt{updating\_a\_pet\_s\_\allowbreak
 status\_should\_be\_observable\_\allowbreak
 when\_retrieving\_that\_pet\_by\_its\_identifier}
 }
 \end{center}
 captures all test scenarios that express this same metamorphic intent, even if the surface wording varies across sessions.

To compare ARMeta with LogiAgent, we report two complementary metrics: (i) operational coverage per session, and (ii) the number of distinct API operation sequences \emph{(Distinct Seq.)} executed by ARMeta, but not found in the LogiAgent-generated scenarios. We also report the corresponding distinct failed sequences (\emph{Distinct Failed}), which include both false positives and true positives, as well as the distinct true positive fault sequences (\emph{Distinct TPFS}) that correspond to real faults. A \emph{sequence} is an ordered list of normalized API operations (method + path). We denote a sequence as $R_1 \rightarrow R_2 \rightarrow R_3$, where each $R_i$ represents the $i$-th request in the sequence. For example, the sequence $R_1 \rightarrow R_2 \rightarrow R_3$ may correspond to \texttt{POST /pet} $\rightarrow$ \texttt{GET /pet/\{petId\}} $\rightarrow$ \texttt{POST /pet/\{petId\}}.

\subsection{Results}
Table~\ref{tab:hlmt-emt} presents the HLMT and EMT generation results over 10 sessions for each case study. On average, ARMeta generates 21.4 HLMTs per session for PetStore in 15.1 minutes and 30.3 HLMTs per session for UserManagement in 22.7 minutes. The higher number of HLMTs for UserManagement is due to its larger number of endpoints and richer API interactions, which allow more metamorphic tests to be derived.

Across all sessions, ARMeta produces 39 semantically unique HLMTs for PetStore and 49 for UserManagement. Compared to the total number of generated HLMTs, this indicates that most tests are reproduced across sessions, while a smaller subset represents new semantic variations. This behavior shows that the approach is reproducible, as similar metamorphic tests are consistently generated in different sessions, while still allowing the discovery of additional test relations over time.

For EMTs, ARMeta generated a total of 211 EMTs for PetStore and 300 EMTs for UserManagement. The lower-than-expected counts are due to syntax errors in the generated EMTs. In PetStore, all three tests triggered \texttt{TypeError}, resulting in failing the Behave scenario. Similarly, in UserManagement, three HLMTs could not be converted into EMTs because they were not fully defined, resulting in \texttt{NotImplementedError}. Despite these conversion failures, ARMeta successfully generated a large number of EMTs, covering HLMTs.
Overall, these results answer RQ1 by demonstrating that the LLM-based approach reliably generates metamorphic tests with high reproducibility and limited but meaningful semantic diversity.

\begin{table*}[t]
\centering
\caption{Per-session HLMT/EMT generation metrics (sessions 1–10 per case study).}
\label{tab:hlmt-emt}
\footnotesize
\begin{tabular}{c|ccccc|c|ccccc}
\toprule
\textbf{Session} &
\multicolumn{5}{c|}{\textbf{PetStore}} &
&
\multicolumn{5}{c}{\textbf{UserManagement}} \\
\cmidrule(r){2-6} \cmidrule(l){8-12}
& Iter. & HLMT & EMTs & Coverage & Time (min)
&& Iter. & HLMTs & EMTs & Coverage & Time (min) \\
\midrule
1  & 9 & 26 & 26 & 100.0\% & 17.95 && 6 & 22 & 22 & 67.9\% & 17.08 \\
2  & 5 & 12 & 12 & 100.0\% & 7.69  && 6 & 22 & 21 & 71.4\% & 17.19 \\
3  & 7 & 21 & 21 & 94.7\%  & 15.90 &&11 & 34 & 34 & 96.4\% & 26.82 \\
4  & 9 & 36 & 36 & 94.7\%  & 24.28 && 6 & 25 & 25 & 67.9\% & 22.83 \\
5  &11 & 30 & 30 & 94.7\%  & 18.28 && 7 & 20 & 20 & 67.9\% & 16.97 \\
6  & 5 & 13 & 11 & 100.0\% & 10.18 && 6 & 24 & 24 & 82.1\% & 14.96 \\
7  & 6 & 16 & 15 & 100.0\% & 10.89 &&13 & 45 & 44 & 96.4\% & 30.21 \\
8  & 7 & 23 & 23 & 100.0\% & 15.42 && 8 & 24 & 24 & 78.6\% & 18.80 \\
9  & 5 & 16 & 16 & 100.0\% & 13.16 &&11 & 48 & 47 & 82.1\% & 30.89 \\
10 & 6 & 21 & 21 & 100.0\% & 17.04 && 9 & 39 & 39 & 82.1\% & 31.00 \\
\midrule
\textbf{Total}
& 68 & 214 & 211 & --- & ---
&& 66 & 303 & 300 & --- & --- \\
\textbf{Average}
& 6.8 & 21.4 & 21.1 & 98.4\% & 13.75 && 6.6& 30.3 & 30.0 & 79.3\% & 15.20 \\
\textbf{Sem. unique} & --- & 39 & --- & --- & --- && --- & 49 & --- & --- & --- \\
\bottomrule
\end{tabular}
\end{table*}

\begin{table*}[t]
\centering
\caption{Per-session results for PetStore and UserManagement showing EMTs, Passed, Failed, TPFT, and TPR (\%). Sessions 1--10 are reported for each case study, along with total and average values.}

\label{tab:emt-af}
\footnotesize
\begin{tabular}{c|ccccc|ccccc}
\toprule
\multirow{2}{*}{\textbf{Session}} &
\multicolumn{5}{c|}{\textbf{PetStore}} &
\multicolumn{5}{c}{\textbf{UserManagement}} \\
\cmidrule(r){2-6} \cmidrule(l){7-11}
& EMTs & Passed & Failed & TPFT & TPR (\%)
& EMTs & Passed & Failed & TPFT & TPR (\%) \\
\midrule
1  & 26 & 14 & 12 & 9  & 75.0  
& 22 & 9  & 13 & 9  & 69.2 \\

2  & 12 & 8  & 4  & 4  & 100.0 
& 21 & 7  & 14 & 7  & 50.0 \\

3  & 21 & 7  & 14 & 9  & 64.3  
& 34 & 9  & 25 & 20 & 80.0 \\

4  & 36 & 17 & 19 & 15 & 78.9  
& 25 & 1  & 24 & 10 & 41.7 \\

5  & 30 & 11 & 19 & 17 & 89.5  
& 20 & 5  & 15 & 7  & 46.7 \\

6  & 11 & 4  & 7  & 6  & 85.7  
& 24 & 6  & 18 & 13 & 72.2 \\

7  & 16 & 8  & 8  & 7  & 87.5  
& 44 & 20 & 24 & 13 & 54.2 \\

8  & 23 & 10 & 13 & 8  & 61.5  
& 24 & 9  & 15 & 9  & 60.0 \\

9  & 15 & 6  & 9  & 8  & 88.9  
& 47 & 17 & 30 & 19 & 63.3 \\

10 & 21 & 9  & 12 & 9  & 75.0  
& 39 & 13 & 26 & 9  & 34.6 \\
\midrule
\textbf{Total}
& 211 & 94 & 117 & 92 & 78.6
& 300 & 96 & 204 & 116 & 56.9 \\

\textbf{Average}
& 21.1 & 9.4 & 11.7 & 9.2 & 80.6
& 30.0 & 9.6 & 20.4 & 11.6 & 57.2 \\
\bottomrule
\end{tabular}
\end{table*}

\begin{table*}[t]
\centering
\caption{Operational coverage (\%), elapsed time (min), and distinct API operations sequence statistics per session for LogiAgent (LA) and ARMeta (AR).}

\label{tab:opcov-time-seq-compact-total}
\footnotesize
\setlength{\tabcolsep}{3pt}
\begin{tabular}{c|cccc|ccc||cccc|ccc}
\toprule
\multirow{2}{*}{\textbf{Session}} 
& \multicolumn{7}{c||}{\textbf{PetStore}} 
& \multicolumn{7}{c}{\textbf{UserManagement}} \\

\cmidrule(r){2-8} \cmidrule(l){9-15}

& LA & LA & AR & AR 
& Distinct & Distinct & Distinct
& LA & LA & AR & AR
& Distinct & Distinct & Distinct \\

& Cov. & Time & Cov. & Time
& Seq. & Failed & TPFS
& Cov. & Time & Cov. & Time
& Seq. & Failed & TPFS \\
\midrule

1  & 73.68 & 50.2 & 100.0 & 18.0 & 9  & 6  & 5 & 60.71 & 50.2 & 67.86 & 17.1 & 10 & 7  & 0 \\
2  & 89.47 & 45.7 & 100.0 & 7.7  & 5  & 3  & 3 & 42.86 & 59.7 & 71.43 & 17.2 & 14 & 12 & 2 \\
3  & 78.95 & 45.1 & 94.74 & 15.9 & 6  & 5  & 5 & 67.86 & 60.1 & 96.43 & 26.8 & 19 & 16 & 8 \\
4  & 84.21 & 55.9 & 94.74 & 24.3 & 9  & 6  & 5 & 46.43 & 55.9 & 67.86 & 22.8 & 13 & 13 & 7 \\
5  & 78.95 & 46.3 & 94.74 & 18.3 & 10 & 8  & 8 & 42.86 & 58.3 & 67.86 & 17.0 & 10 & 8  & 1 \\
6  & 84.21 & 48.1 & 100.0 & 10.2 & 7  & 5  & 4 & 42.86 & 58.1 & 82.14 & 15.0 & 17 & 12 & 5 \\
7  & 84.21 & 48.3 & 100.0 & 10.9 & 6  & 3  & 3 & 82.14 & 58.3 & 96.43 & 30.2 & 32 & 20 & 4 \\
8  & 73.68 & 49.6 & 100.0 & 15.4 & 12 & 7  & 4 & 17.86 & 59.6 & 78.57 & 18.8 & 15 & 12 & 4 \\
9  & 84.21 & 49.9 & 100.0 & 13.2 & 5  & 3  & 3 & 42.86 & 52.8 & 82.14 & 30.9 & 21 & 17 & 7 \\
10 & 84.21 & 47.9 & 100.0 & 17.0 & 9  & 5  & 5 & 21.43 & 57.9 & 82.14 & 31.0 & 16 & 11 & 2 \\
\midrule
\textbf{Mean} 
& 81.6 & 48.7 & 98.4 & 15.1 
& 7.8 & 5.1 & 4.5
& 46.8 & 57.1 & 79.3 & 22.7
& 16.7 & 12.8 & 4.0 \\
\midrule
\textbf{Total} 
& -- & -- & -- & -- 
& 78 & 51 & 45
& -- & -- & -- & -- 
& 167 & 128 & 40 \\
\midrule
\textbf{Overall Cov.} 
& \multicolumn{2}{c}{18/19} & \multicolumn{2}{c}{19/19} & \multicolumn{3}{c||}{--}
& \multicolumn{2}{c}{25/28} & \multicolumn{2}{c}{28/28} & \multicolumn{3}{c}{--} \\
\bottomrule
\end{tabular}
\end{table*}

Table \ref{tab:emt-af} reports fault detection results per session using EMTs for both case studies. EMTs are tests that are executed correctly. For each session, the table shows the number of generated EMTs, passed and failed tests, TPFTs, and the TPR.

For PetStore, ARMeta generated an average of 21.1 EMTs per session. Out of 117 failed EMTs across sessions, 92 were TPFTs, resulting in an overall TPR of 78.6\%. This shows that most failures correspond to real faults. For UserManagement, ARMeta produced an average of 30.0 EMTs per session with 116 out of 204 failed EMTs being TPFTs, giving an overall TPR of 56.9\%.  The lower TPR is mainly attributed to a higher number of \emph{false positives}. The API is stricter and requires additional setup (e.g., valid usernames and required fields), which leads to failures caused by incorrect test assumptions rather than actual SUT faults. The results may also be influenced by the prompt construction, particularly the use of few-shot examples from PetStore.
Across the TPFT, root causes broadly fall into: (i) \emph{request contract violations} (e.g., missing required fields leading to \texttt{400} errors), 
(ii) \emph{response contract violations} (e.g., invalid date-time in headers or unexpected response structure), 
(iii) \emph{server crashes} (5xx responses during valid sequences), and 
(iv) \emph{timeouts} (no response within the expected time due to API instability).

Overall, the table answers RQ2 and shows that ARMeta can effectively generate executable fault-revealing EMTs. However, the TPR varies across systems. This indicates the need to reduce false positives and improve stability in future work.

Table~\ref{tab:opcov-time-seq-compact-total} shows that LogiAgent and ARMeta differ not only in coverage but also in the API operation sequences they execute. LogiAgent generates linear scenario scripts directly from its test generator. In contrast, ARMeta constructs the sequence of its interactions based on the sequence of seed and follow-up requests. Although both approaches execute APIs in a linear order, ARMeta produces different API sequences compared to LogiAgent, which leads to the exploration of different execution paths. As a result, ARMeta exercises request sequences that LogiAgent does not explore.

For instance, ARMeta executes an API request sequence:
    \begin{itemize}
      \item \texttt{POST /user}
        \item  \texttt{GET /user/login}
          \item \texttt{GET /user/login}
    \end{itemize}
    The MR here is \emph{repeat-login consistency}: after creating a user, two consecutive logins with the same credentials should both succeed and return responses with the same structure (including valid \texttt{X-Rate-Limit} and \texttt{X-Expires-After} headers). The test fails because \texttt{X-Expires-After} is not a valid date-time, i.e., a response-contract violation.
, which LogiAgent does not generate; this test fails because the API returns an invalid \texttt{X-Expires-After} date-time (a response-contract violation).
Likewise, ARMeta executes the following APIs:
\begin{itemize}
 \item \texttt{POST /pet} 
  \item\texttt{GET /pet/\{petId\}} 
   \item\texttt{GET /pet/\{petId\}} 
    \item\texttt{POST /pet/\{petId\}}
     \end{itemize} 
     
The test fails because the API does not update the pet status (state inconsistency) as discussed in Figure \ref{fig:mr-petstore-subfig}. These failures are not discovered by LogiAgent because the metamorphic request ordering exercised by ARMeta is absent from LogiAgent’s generated scenarios. 

Looking at all generated tests executing API sequences across the ten sessions, ARMeta consistently outperformed LogiAgent in both coverage and detected those faults that were not detected by LogiAgent, as shown in Table  \ref{tab:opcov-time-seq-compact-total}. For PetStore, ARMeta achieved near-complete operational coverage while LogiAgent missed some operations, and ARMeta generated 51 distinct failed API sequences, of which 45 were true positives failing sequences, none of which were found by LogiAgent. For UserManagement, ARMeta reached full coverage, whereas LogiAgent covered fewer operations, and ARMeta produced 128 failing API sequences with 40 true positives, again with no overlap with LogiAgent’s results. Overall, these results show that ARMeta covers more system behavior by generating API call sequences that exercise different execution paths and reveal failures not detected by LogiAgent, even when LogiAgent is given similar or longer execution time. Therefore, with respect to RQ3, the findings indicate that ARMeta can complement existing scenario-based API testing approaches by revealing additional behaviors and faults rather than replacing them.

The replication package for the evaluation above can be found at \cite{ARMeta-v1-672B}.

\section{Limitations and Future Work}
\label{sec:limitations}
During repeated sessions of ARMeta, we observed that semantically equivalent HLMTs were sometimes implemented as different EMTs across sessions. Although the workflow is sequential within a single execution, the EMT generation step may select different specification-valid APIs when implementing the same HLMT across executions. While this variability can help reveal faults, it also indicates instability in the transformation process. Future work will focus on refining the LLM prompts and improving determinism in EMT generation.

In some cases, an EMT using one API executed successfully, whereas another EMT generated for the same HLMT but using a functionally equivalent API exposed a fault, such as a contract violation or server error. This demonstrates that ARMeta can uncover inconsistencies between APIs intended to provide similar functionality. However, this behavior is emergent rather than explicitly designed in the workflow. We plan to systematically incorporate this cross-API consistency analysis into the framework.

Our results also show that some HLMTs could not be successfully transformed into executable EMTs, even with the refinement layer. Failures were primarily caused by TypeErrors, incomplete code generation, or incorrect parameter handling. As future work, we plan to integrate more advanced code-specialized LLMs (e.g., GPT-5.2 Codex or Anthropic’s Claude Opus) to improve the robustness and reliability of the transformation process and enhance the quality of the generated test artifacts.

Both ARMeta and LogiAgent rely on LLMs, which introduce inherent nondeterminism. Although we reduce randomness by using temperature~0 and a fixed seed, identical inputs may still produce different metamorphic relations or test implementations across sessions. Consequently, the generated scenarios and observed failures may vary. Furthermore, ARMeta depends on OpenAPI specifications as the source of truth for endpoints and parameters. If the specification is incomplete, outdated, or inaccurate, the framework may generate invalid tests or fail to detect undocumented behaviors. Addressing specification quality and incorporating specification validation mechanisms represent important directions for future research.

\section{Conclusion}
In this paper, we discussed that metamorphic testing motivates a useful structure for REST APIs: execute a seed call (or short sequence), execute a transformed follow-up, and check a metamorphic relation between their outcomes. When the relation assertion is reachable, it provides an oracle based on consistency across executions. In practice, many REST API tests fail earlier due to server crashes, exceptions, and specification mismatches. These failures are still valuable because they expose robustness and contract problems that block higher-level checks.

ARMeta uses a multi-agent workflow to generate metamorphic relations, turn them into Given–When–Then scenarios, and then produce executable Behave test code. We evaluated ARMeta on two real APIs and found that it can generate executable tests and reach good operation coverage. 

Overall, ARMeta shows that metamorphic, transformation-based testing can complement existing API testing approaches by exploring more behaviors and helping reveal robustness and mismatches between documentation and API implementation problems. In future work, we plan to use code-specific LLMs make failure reports easier to understand.

\section*{Acknowledgment}
This work was funded by the Business
Finland via the Virtual Sea Trial project (VST), under grant 7187/31/2023 and Finnish Ministry of Educa-
tion and Culture’s Doctoral Education Pilot under Decision
No. VN/3137/2024-OKM-6 (The Finnish Doctoral Program
Network in Artificial Intelligence, AI-DOC).


\end{document}